\documentclass[12pt]{article}

\usepackage{amsfonts,amscd,amssymb,amsmath}
\usepackage{verbatim}

\date{}

\usepackage{mathptmx}       % selects Times Roman as basic font
\usepackage{helvet}         % selects Helvetica as sans-serif font
\usepackage{courier}        % selects Courier as typewriter font
\usepackage{type1cm}        % activate if the above 3 fonts are
\usepackage{makeidx}         % allows index generation
\usepackage{graphicx}        % standard LaTeX graphics tool
\usepackage{multicol}        % used for the two-column index
\usepackage[bottom]{footmisc}% places footnotes at page bottom

\begin{document}

\title{Towards  $p$-Adic Matter in the Universe}
% Use \titlerunning{Short Title} for an abbreviated version of
% your contribution title if the original one is too long
%\author{Branko Dragovich}
%\authorrunning{Dragovich}
% your contribution title if the original one is too long
\author{Branko Dragovich\thanks{dragovich@ipb.ac.rs} \\ Institute of Physics, University of
Belgrade \\ Pregrevica 118, 11080 Zemun, Belgrade, Serbia }
%\and Name of Second Author \at Name, Address of Institute \email{name@email.address}}
\maketitle

\abstract{~Starting from $p$-adic string theory with tachyons,   we
introduce a new kind of non-tachyonic  matter which may play an
important role in evolution of the Universe. This matter retains
nonlocal and nonlinear $p$-adic string dynamics, but does not suffer
of negative square mass. In space-time dimensions $D = 2 + 4k$, what
includes $D = 6,\, 10,\, ...,\, 26,$ the kinetic energy term also
maintains correct sign. In these spaces this $p$-adic matter
provides negative cosmological constant and time-dependent scalar
field solution with negative potential. Their possible cosmological
role is discussed. We have also connected non-locality with string
world-sheet in effective Lagrangian for $p$-adic string.}

\section{Introduction}
\label{sec:1}

Observational modern cosmology has achieved significant progress
by high precision experiments in the last decades.  One of the
greatest cosmological achievements was discovery of accelerated
expansion of the Universe in 1998. If General Relativity is theory
of gravity for the Universe as a whole then about $96 \%$ of its
energy content is of unknown nature. This dark side of the
Universe consists of about $23 \%$ of  \textit{dark matter} and
$73 \%$ of \textit{dark energy}. Dark matter is supposed to be
responsible for anomaly large rotational velocities in the spiral
galaxies. Dark energy has negative pressure and should govern the
Universe accelerated expansion (as a recent review see
\cite{kinezi}). Thus, according to this point of view, there is
now only about $4 \%$ of visible matter which is described by the
Standard Model of particle physics. Although dark matter and dark
energy are well adopted among majority of scientists, they are not
directly verified in the laboratory and still remain hypothetical
forms of matter. Also General Relativity has not been so far
confirmed at the cosmic scale. For these reasons, there is not yet
commonly accepted theoretical explanation of the Universe
acceleration. This situation has influenced  also alternative
approaches, mostly related to modification of General Relativity
(a recent review in \cite{clifton}).

While observational cosmology is in an arising state, theoretical
cosmology is facing a big challenge. An exotic matter and
modification of gravity are two alternative theoretical
approaches. Since string theory is the best candidate for
unification of matter elementary constituents and fundamental
interactions, some theoretical ideas come from string theory and
we consider the following one. According to the adelic product
formula for scalar string amplitudes it follows that $p$-adic
strings are at equal footing with ordinary strings (reviews on
$p$-adic strings and adelic product formula can be found in
\cite{freund,vladimirov}). Hence, if visible matter is composed of
ordinary strings then there should be some matter made of $p$-adic
strings. It is natural to assume that dark side of the Universe
contains some kinds of $p$-adic matter. In $p$-adic string theory
world-sheet has non-Archimedean (ultrametric) geometry and it
should also somehow modify gravity. It is feasible that future
theoretical cosmology will be a result of both modification of
General Relativity and inclusion of new kinds of matter.

Inspired by $p$-adic string theory  it has been already
investigated some nonlocal  modifications of General Relativity
(see, e.g. \cite{dragovich1} and references therein) and
cosmology, see, e.g. \cite{arefeva01,koshelev1} and references
therein. In this article we consider some modification of open
scalar $p$-adic strings as candidates for a new kind of matter in
the Universe. In Section \ref{sec:2} we present various aspects of
$p$-adic string theory necessary for comprehensive exposition. It
also contains introduction of non-tachyonic $p$-adic matter.
Section \ref{sec:3} is related to some adelic approaches to
cosmology.

\section{$p$-Adic Strings}
\label{sec:2}

$p$-Adic strings are introduced in 1987 by Volovich in his paper
\cite{volovich1}. $p$-Adic string theory is mainly related to
strings which have only world-sheet $p$-adic and all other their
properties are the same with theory of ordinary strings
\cite{freund1}. Having exact Lagrangian at the tree level,
$p$-adic scalar strings have attracted significant attention in
string theory and nonlocal cosmology. To be more comprehensive and
self consistent we shall first give a brief review of $p$-adic
numbers and their applicability in modern mathematical physics.

\subsection{$p$-Adic Numbers and Their Applicability}

$p$-Adic numbers are discovered by Kurt Hansel in 1897 as a new
tool in number theory. In modern approach to introduce $p$-adic
numbers one usually starts with the field  $\mathbb{Q}$ of
rational numbers. Recall that according to the Ostrowski theorem
any non-trivial norm on $\mathbb{Q}$ is equivalent either to the
usual absolute value $|\cdot|_\infty$ or to a $p$-adic norm
($p$-adic absolute value) $|\cdot|_p$. A rational number $x =
p^\nu \, \frac{a}{b}$, where integers $a$ and $b \neq 0$ are not
divisible by prime number $p$, by definition has $p$-adic norm
$|x|_p = p^{-\nu}$ and $|0|_p = 0 .$  Since $|x + y|_p \leq
\mbox{max} \{|x|_p \,, |y|_p \}$,  $p$-adic norm is a
non-Archimedean (ultrametric) one. As completion of $\mathbb{Q}$
with respect to the absolute value $|\cdot|_\infty$ gives the
field $\mathbb{Q}_\infty \equiv \mathbb{R}$ of real numbers, by
the same procedure using $p$-adic norm $|\cdot|_p$ one gets the
field $\mathbb{Q}_p\,\,$ of $p$-adic numbers (for any prime number
$p = 2,\, 3\,, 5\, \cdots$). Any number $0 \neq x \in
\mathbb{Q}_p\,\,$ has its unique canonical representation
\begin{align} \label{equ.1}
x = p^{\nu} \, \sum_{n = 0}^{+ \infty} \, x_n\, p^n  \,, \quad \nu
 \in \mathbb{Z}\,, \quad x_n \in \{0,\, 1,\, \cdots, \, p-1 \},
\quad x_0 \neq  0 .
\end{align}
 $\mathbb{Q}_p$ is locally compact, complete and totally
disconnected topological space. There is a rich structure of
algebraic extensions of $\mathbb{Q}_p$.

There are many possibilities for mappings between $\mathbb{Q}_p$.
The most elaborated is analysis related to mappings $\mathbb{Q}_p
 \to \mathbb{Q}_p$ and $\mathbb{Q}_p \to \mathbb{C}$. Usual complex valued
 functions of $p$-adic argument are additive $\chi_p(x) =  e^{2\pi i \{x \}_p}$ and multiplicative
 $|x|^s$ characters, where $\{x \}_p$ is fractional part of $x$ and $s \in \mathbb{C}$ (for many aspects of
 $p$-adic numbers and their analysis, we refer to \cite{freund,vladimirov,gelfand,schikhof}).

The field $\mathbb{Q}$ of rational numbers, which is dense in
$\mathbb{Q}_p$ and $\mathbb{R}$, is also important in physics. All
values of measurements are rational numbers. Any measurement  is
comparison of two quantities of the same kind and it is in close
connection with the Archimedean axiom. Set of rational numbers
obtained in the process of repetition of measurement of the same
quantity is naturally provided by usual absolute value. Hence,
results of measurements are not rational numbers with $p$-adic
norm but with real norm. It means that measurements give us real
and not $p$-adic rational numbers. Then the following question
arises: Being not results of measurements, what role $p$-adic
numbers can play in description of something related to physical
reality? Recall that we already have similar situation with
complex numbers, which are not result of direct measurements but
they are very useful. For example,  in quantum mechanics wave
function is basic theoretical tool which contains all information
about quantum system but cannot be directly measured  in
experiments. $p$-Adic numbers should play unavoidable role where
application of real numbers is inadequate. In physical systems
such situation is at the Planck scale, because it is not possible
to measure distances smaller than the Planck length.  It should be
also the case with very complex phenomena of living and cognitive
systems. Thus, we expect inevitability  of $p$-adic numbers at
some deeper level in understanding  of the content, structure and
evolution of the Universe in its parts as well as a whole. The
first steps towards  probe of $p$-adic levels of knowledge is
invention of relevant mathematical methods and construction of the
corresponding physical models. A brief overview of $p$-adic
mathematical physics is presented in \cite{dragovich2}. It
includes both $p$-adic valued and real (complex) valued functions
of $p$-adic argument. In the sequel we shall consider $p$-adic
strings, non-tachyonic $p$-adic matter and its some possible  role
in modern cosmology.

\subsection{Scattering Amplitudes and Lagrangian for Open Scalar $p$-Adic Strings }

Like ordinary strings, $p$-adic strings are introduced by
construction of their scattering amplitudes. The simplest
amplitude is for scattering of two open scalar strings. Recall the
 crossing symmetric Veneziano amplitude for ordinary strings
\begin{align} \label{equ.2}
 A_\infty (a, b) = g_\infty^2 \,\int_{\mathbb{R}}
|x|_\infty^{a-1}\, |1 -x|_\infty^{b-1}\, d_\infty x   = g_\infty^2
\, \frac{\zeta (1 - a)}{\zeta (a)}\, \frac{\zeta (1 - b)}{\zeta
(b)}\, \frac{\zeta (1 - c)}{\zeta (c)},
\end{align}
where $a+b+c = 1.$ The crossing symmetric Veneziano amplitude for
scattering of two open scalar $p$-adic strings is direct analog of
(\ref{equ.2}) \cite{freund1}, i. e.
\begin{align} \label{equ.3}
A_p (a, b) = g_p^2 \, \int_{\mathbb{Q}_p} |x|_p^{a-1}\, |1
-x|_p^{b-1}\, d_p x   = g_p^2 \, \frac{1 - p^{a - 1}}{1 -
p^{-a}}\, \frac{1 - p^{b - 1}}{1 - p^{-b}}\, \frac{1 - p^{c -
1}}{1 - p^{-c}}.
\end{align}
Integral expressions in (\ref{equ.2}) and (\ref{equ.3}) are the
Gel'fand-Graev-Tate beta functions on $\mathbb{R}$ and
$\mathbb{Q}_p$, respectively \cite{gelfand}. Note that here, by
definition, ordinary and $p$-adic strings differ only in
description of their world-sheets -- world-sheet of $p$-adic
strings is presented by $p$-adic numbers.  Kinematical variables
contained in $a, b, c$ are the same real numbers in both cases. It
is worth noting that the final form of  Veneziano amplitude for
$p$-adic strings (\ref{equ.3}) is rather simple and presented by
an elementary function.

It is remarkable that there is an effective field description of
the above  open  $p$-adic strings. The corresponding Lagrangian is
very simple and  at the tree level describes not only four-point
scattering amplitude but also all higher (Koba-Nielsen) ones. The
exact form of this  Lagrangian for effective scalar field
$\varphi,$ which describes open $p$-adic string tachyon, is
\begin{align} \label{equ.4}
 {\cal L}_p = \frac{m^D}{g^2}\, \frac{p^2}{p-1} \Big[
-\frac{1}{2}\, \varphi \, p^{-\frac{\Box}{2 m^2}} \, \varphi +
\frac{1}{p+1}\, \varphi^{p+1} \Big],
\end{align}
where $p$
 is any prime number, $D$ -- spacetime dimensionality, $\Box = - \partial_t^2  + \nabla^2$ is the
$D$-dimensional d'Alembertian and  metric has signature $(- \, +
\, ...\, +)$ \cite{freund2,frampton}. This is nonlocal and
nonlinear Lagrangian. Nonlocality is in the form of infinite
number of spacetime derivatives
\begin{align} \label{equ.5}
  p^{-\frac{\Box}{2 m^2}} = \exp{\Big( - \frac{\ln{p}}{2 m^2}\,
\Box \Big)} = \sum_{k \geq 0} \, \Big(-\frac{\ln p}{2 m^2} \Big)^k
\, \frac{1}{k !}\, \Box^k
\end{align}
and it is a consequence of the fact that strings are extended
objects.

The corresponding potential $\mathcal{V}(\varphi)$ for Lagrangian
(\ref{equ.4}) is $\mathcal{V}_p(\varphi) = - \mathcal{L}_p (\Box =
0)$, which  the explicit form is
\begin{align} \label{equ.4V}
 \mathcal{V}_p(\varphi) = \frac{m^D}{g^2}\, \Big[
\frac{1}{2} \frac{p^2}{p-1}\, \varphi^2  \, - \,
\frac{p^2}{p^2-1}\, \varphi^{p+1} \Big].
\end{align}
It has local minimum  $\mathcal{V}_p (0) = 0.$ If $p \neq 2$ there
are two local maxima at $\varphi =\pm 1$ and there is one local
maximum $\varphi = + 1$ when $p = 2.$

\begin{figure}[t]
\begin{center}
%\sidecaption
% Use the relevant command for your figure-insertion program
% to insert the figure file.
% For example, with the graphicx style use
\includegraphics[scale=.60]{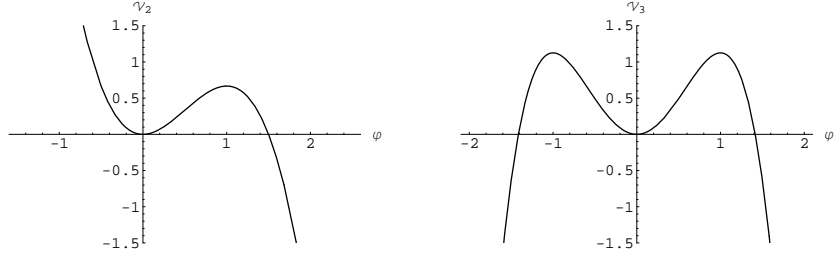}
% If no graphics program available, insert a blank space i.e. use
%\picplace{5cm}{2cm} % Give the correct figure height and width in cm
\caption{The $2$-adic string potential $\mathcal{V}_2 (\varphi)$
(on the left) and  $3$-adic potential $\mathcal{V}_3 (\varphi)$
(on the right) of standard Lagrangian (\ref{equ.4}), where
potential is presented by expression (\ref{equ.4V}) with
$\frac{m^D}{g^2} =1$.}
%\texttt{sidecapion} command to flush the caption on the left side
%of the page. If the figure is positioned at the top of the page,
%align the sidecaption with the top of the figure -- to achieve
%this you simply need to use the optional argument \texttt{[t]}
%with the \texttt{sidecaption} command}
\end{center}
\label{fig:1}       % Give a unique label
\end{figure}

The equation of motion for the field $\varphi$ is
\begin{align}\label{equ.6}
 p^{-\frac{\Box}{2 m^2}}\, \varphi = \varphi^p
\end{align}
and it has trivial solutions $\varphi = 0$ and $\varphi =1$, and
another $\varphi = -1$ for $p \neq 2$. There are also
inhomogeneous solutions in any direction $x^i$ resembling solitons
\begin{align}\label{equ.7}
 \varphi (x^i) = p^{\frac{1}{2(p-1)}}\, \exp \Big( - \frac{p-1}{2
 \, p \ln p}\, m^2\, (x^i)^2 \Big).
\end{align}
There is also homogeneous and isotropic time-dependent solution
\begin{align}\label{equ.8}
 \varphi (t) = p^{\frac{1}{2(p-1)}}\, \exp \Big(  \frac{p-1}{2
 \, p \ln p}\, m^2 \, t^2 \Big).
\end{align}
Solution (\ref{equ.8}) (and analogously (\ref{equ.7})) can be
obtained using identity
\begin{align}
e^{A\, \partial_t^2} \, e^{B\, t^2} = \frac{1}{\sqrt{1 - 4 A B}}\,
e^{\frac{B\, t^2}{1 - 4 A B}},    \quad 1- 4 A B > 0.
\end{align}
Note that the sign of the above field solutions $\varphi (x^i)$
and $\varphi (t)$ can be also minus $(-)$ when $p \neq 2$. Various
aspects of $p$-adic string theory with the above effective field
have been pushed forward  by papers  \cite{sen,moeller}.

It is worth noting that Lagrangian (\ref{equ.4}) is written
completely in terms of real numbers and there is no explicit
dependence on the world sheet. However, (\ref{equ.8}) can be
rewritten in the following form:
\begin{align} \nonumber
 {\cal L}_p = &\frac{m^D}{g^2}\, \frac{p^2}{p-1} \Big[
\frac{1}{2}\, \varphi \, \int_{\mathbb{R}}
\Big(\int_{\mathbb{Q}_p\setminus \mathbb{Z}_p} \chi_p (u)
|u|_p^{\frac{k^2}{2m^2}} du \Big) \tilde{\varphi}(k)\, \chi_\infty
(kx)\, d^4k  \\ &+ \frac{1}{p+1}\, \varphi^{p+1} \Big],
\label{equ.10}
\end{align}
where $\chi_\infty (kx) = e^{-2\pi i kx}$ is the real additive
character. Since $\int_{\mathbb{Q}_p} \chi_p(u) |u|^{s-1} du =
\frac{1 - p^{s-1}}{1-p^{-s}} = \Gamma_p (s)$ and it is present in
the scattering amplitude (\ref{equ.3}), one can say that
expression $\int_{\mathbb{Q}_p\setminus \mathbb{Z}_p} \chi_p (u)
|u|_p^{\frac{k^2}{2m^2}} du$ in (\ref{equ.10}) is related to the
$p$-adic string world-sheet.

\subsection{New Kind of Matter, Which Has  Origin in $p$-Adic Strings}

Since there are infinitely many primes $p,$ in principle it can be
infinitely many kinds of $p$-adic strings. We suppose that for all
but a finite set $\mathcal{P}$ of primes $p$ these $p$-adic
strings are in their  local potential minimum, i.e. $\varphi_p =
0,$ and consequently $\mathcal{L}_p = 0,$ for all $p \notin
\mathcal{P}.$ This can be a result of tachyon condensation.
Further we suppose that in the remaining  finite set of strings
there was a transition $m^2 \to - m^2$ (transition from tachyon to
no-tachyon  state), what could be a result of some quantum effect
which was happened before process of tachyon condensation was
finished. For simplicity, we shall assume two kinds of such
strings and denote their set by $\mathcal{P} = \{q, \,  \ell\}.$
In the sequel for these strings in the above expressions
(\ref{equ.4}) - (\ref{equ.10}) one has to replace $m^2$ by $-m^2$
and $m^D \to (-1)^{\frac{D}{2}} m^D ,$ where spacetime
dimensionality $D$ is even. Note that there is transition $m^D \to
- m^D$ for critical dimensions $D=26$ and $D= 10,$ but for $D =4$
there is no change of sign. The corresponding potentials for $p=2$
and $p=3,$ and $D= 2 + 4k ,$ are presented at  Figure 2. To make
distinction with tachyons we denote these new scalar strings by
$\phi_p, \, p \in \mathcal{P}.$

The related new Lagrangian is
\begin{align} \label{equ.4n}
 {L}_p  = (-1)^{\frac{D}{2}} \, \frac{m^D}{g^2}\, \frac{p^2}{p-1} \Big[
-\frac{1}{2}\, \phi \, p^{\frac{\Box}{2 m^2}} \, \phi +
\frac{1}{p+1}\, \phi^{p+1} \Big]
\end{align}
with the corresponding potential
\begin{align} \label{equ.4Vn}
 {V}_p(\phi) =  (-1)^{\frac{D}{2}} \, \mathcal{V}_p (\phi)  =  (-1)^{\frac{D}{2}} \, \frac{m^D}{g^2}\, \Big[
\frac{1}{2} \frac{p^2}{p-1}\, \phi^2  \, - \, \frac{p^2}{p^2-1}\,
\phi^{p+1} \Big].
\end{align}

The equation of motion for scalar strings $\phi_p$ is
\begin{align}\label{equ.6n}
 p^{\frac{\Box}{2 m^2}}\, \phi_p = \phi_p^p
\end{align}
and it has trivial solutions $\phi_p = 0$ and $\phi_p =1$, and
another $\phi_p = -1$ for $p \neq 2$. What is local maximum and
minimum depends on dimensionality $D$. When $D = 2 + 4k$, solution
$\phi_p =0$ is a local maximum, and $\phi_2 = +1$ and $\phi_p =
\pm 1, \, p\neq 2$ are local minima. For any dimensionality,
nontrivial solutions become now
\begin{align}\label{equ.7n}
 \phi_p (x^i) = p^{\frac{1}{2(p-1)}}\, \exp \Big( \frac{p-1}{2
 \, p \ln p}\, m^2\, (x^i)^2 \Big),
\end{align}
\begin{align}\label{equ.8n}
 \phi_p (t) = p^{\frac{1}{2(p-1)}}\, \exp \Big( - \frac{p-1}{2
 \, p \ln p}\, m^2\, t^2 \Big).
\end{align}
$D$-dimensional solution of (\ref{equ.6n}) is product of solutions
(\ref{equ.7n}) and (\ref{equ.8n})  (see also \cite{vladimirov1}),
i.e.
\begin{align}\label{equ.9n}
 \phi_p (x) = p^{\frac{D}{2(p-1)}}\, \exp \Big( \frac{p-1}{2
 \, p \ln p}\, m^2\, x^2 \Big), \quad x^2 = -t^2 + \sum_{i=1}^{D-1} x_i^2 .
\end{align}

 Lagrangian for this collection of two distinct strings
$\{ \phi_q , \, \phi_{\ell} \}$  is of the form
\begin{align}
{L}_{\mathcal{P}} = \sum_{p \in \mathcal{P}}\,  { L}_p = \sum_{p
\in \mathcal{P}} (-1)^{\frac{D}{2}} \, \frac{m^D}{g^2}\,
\frac{p^2}{p-1} \Big[ - \frac{1}{2}\, \phi_p \, p^{\frac{\Box}{2
m^2}} \, \phi_p + \frac{1}{p+1}\, \phi_p^{p+1} \Big].
\label{equ.11}
\end{align}
String field $\phi_{\ell}$ in (\ref{equ.11}) we shall  consider
in its vacuum state $\phi_{\ell} = + 1$ or $- 1$ with  Lagrangian
\begin{align} \label{equ.12}
L_{\ell} = - V_{\ell}(\pm 1)= (-1)^{\frac{D}{2}+1} \,
\frac{m^D}{g^2}\, \frac{\ell^2}{2(\ell +1)} \sim - \Lambda
\end{align}
related to the cosmological constant $\Lambda$ (prime index $\ell$
may remind $\Lambda$).
 Note that vacuum state $\phi_{\ell} = \pm 1$ is stable only
in spaces with dimension $D = 2 + 4k$.

Field $\phi_q$  corresponds to time-dependent solution
(\ref{equ.8n}) in dimensions which satisfy  $(-1)^\frac{D}{2} =
-1,$ and the form of the corresponding potential is presented at
Figure 2. As a simple example, one can take $D = 6$ as respective
solution of condition $(-1)^\frac{D}{2} = -1.$ For the case $D=
6,$ or any other $D = 2 +4k$, we have ($q$ may remind
quintessence)
\begin{align}\label{equ.8n1}
 \phi_q (t) = q^{\frac{1}{2(q-1)}}\, \exp \Big( - \frac{q-1}{2
 \, q \ln q}\, m^2\, t^2 \Big)
\end{align}
which corresponds to  potential of the form at  Figure 2. At the
moment $t = 0$ (the big bang) the field $\phi_q$ has its maximum
which is $\phi_q (0) = q^{\frac{1}{2(q-1)}}$ and it is  a bit
larger than $1$. Then by increasing of time  $\phi_q (t)$ is
decreasing and $\phi_q (t) \to 0$ as $t \to +\infty.$ The
situation is symmetric with respect to transformation $t \to - t.$

If we consider $\phi_q (t)$ in spaces of dimension $D = 4 k$, and
in particular $D =4,$ then we  face by two problems. First, the
kinetic energy term has not correct sign. Second, the position of
field at moment $t = 0$ is $\phi_q (0) = q^{\frac{1}{2(q-1)}} > 1$
and it should have rolling to $-\infty,$ instead of to $0$, what
contradicts to the time dependence (\ref{equ.8n1}) of the field
(see also Fig. 1).

 %The corresponding energy density for  solution
%(\ref{equ.8n1}) is analog of energy for soliton solution in
%\cite{yang}, and it is
%\begin{align} \nonumber
%\rho_q (t) = & - \frac{m^{D+1}}{g^2}\, \frac{q-1}{q+1}
%\sqrt{\frac{2\pi}{(q^2 -1)\ln{q}}}  \, q^{\frac{q}{q-1}}\, |t| \,
%\mbox{erf}\Big[ \frac{q-1}{q+1}\, \sqrt{\frac{q^2 -1}{2 q \ln q}}
%m
%|t|\Big] \\
%& \times\exp{\Big[- \frac{2(q-1)}{ (q +1)\ln q} m^2 t^2} \Big].
%\label{equ.13}
%\end{align}
%The solution  $\phi_3 (t)\, $ (\ref{equ.8n1}) and related energy
%density $\rho_3(t)$ (\ref{equ.13}) are illustrated at Fig. 3.

%In $D =6$, there can exist solution ($d$ is to remind darkness)
%\begin{align}\label{equ.8n2}
% \phi_d (t) = d^{\frac{1}{2(d-1)}}\, \exp \Big( - \frac{d-1}{2
%m^2 \, d \ln d}\, t^2 \Big). \end{align} Time evolution of $\phi_d
%(t)$ is similar to the time behavior of $\phi_q (t)$. Main
%distinction between effects of fields $\phi_q (t)$ and $\phi_d
%(t)$ is in their potentials -- $V(\phi_q)$ is non-negative and
%$V(\phi_d)$ is non-positive (cf. Fig. 1 and Fig. 2).

\begin{figure}[t]
\begin{center}
%\sidecaption
% Use the relevant command for your figure-insertion program
% to insert the figure file.
% For example, with the graphicx style use
\includegraphics[scale=.45]{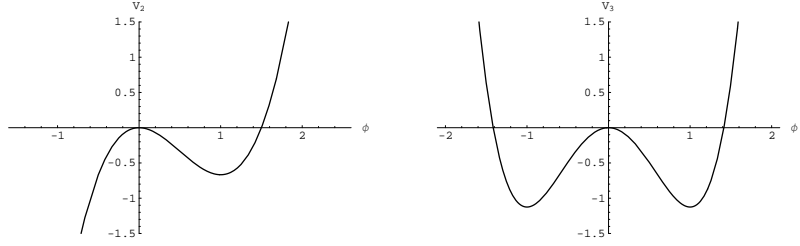}
% If no graphics program available, insert a blank space i.e. use
%\picplace{5cm}{2cm} % Give the correct figure height and width in cm
\caption{New potentials $V_2(\phi)$ and $V_3(\phi)$, which are
related to Lagrangian (\ref{equ.4n}) with dimensionality $D$
satisfying  condition $(-1)^{\frac{D}{2}} = -1,$ i.e. $D = 2 +4k,$
and $\frac{m^D}{g^2} =1$.}
%\texttt{sidecapion} command to flush the caption on the left side
%of the page. If the figure is positioned at the top of the page,
%align the sidecaption with the top of the figure -- to achieve
%this you simply need to use the optional argument \texttt{[t]}
%with the \texttt{sidecaption} command}
\end{center}
\label{fig:1}       % Give a unique label
\end{figure}

%\subsection{Case of all primes taken together}

%\begin{figure}[t]
%\begin{center}
%\sidecaption
% Use the relevant command for your figure-insertion program
% to insert the figure file.
% For example, with the graphicx style use
%\includegraphics[scale=.52]{slika3.eps}
% If no graphics program available, insert a blank space i.e. use
%\picplace{5cm}{2cm} % Give the correct figure height and width in cm
%\caption{Time-dependent field $\phi_3(t)$ and the corresponding
%energy $\rho_3(t)$ related to Lagrangian (\ref{equ.4n}) in
%space-time of dimensionality  $D = 2 +4k$ and ${m^2}= {g^2} =1$.}
%\texttt{sidecapion} command to flush the caption on the left side
%of the page. If the figure is positioned at the top of the page,
%align the sidecaption with the top of the figure -- to achieve
%this you simply need to use the optional argument \texttt{[t]}
%with the \texttt{sidecaption} command}
%\end{center}
%\label{fig:1}       % Give a unique label
%\end{figure}

\section{Adelic Cosmological Modelling}
\label{sec:3}

In the preceding section we have seen that the fields of $p$-adic
and real numbers can be obtained by completion of the field of
rational numbers, and that $\mathbb{Q}$ is dense in $\mathbb{Q}_p$
as well as in $\mathbb{R}.$ This gives rise to think that it
should exist some way for unification of $p$-adic and real
numbers. A unified and simultaneous treatment of $p$-adic and real
numbers is through concept of adeles.  Adelic formalism is a
mathematical method how to connect $p$-adic with ordinary real
models.

\subsection{Adeles and Their Applicability}

An adele $\alpha$ is an infinite sequence made od real and
$p$-adic numbers in the form
\begin{align} \label{equ.3.1}
\alpha = (\alpha_\infty ,\, \alpha_2 ,\, \alpha_3, \, \cdots ,\,
\alpha_p \,,\, \cdots) \,,           \quad \alpha_\infty \in
\mathbb{R} \,, \,\, \alpha_p \in \mathbb{Q}_p \,,
\end{align}
where for all but a finite set $\mathcal{P}$ of primes $p$ it has
to be $\alpha_p \in \mathbb{Z}_p = \{ x\in \mathbb{Q}_p \, : |x|_p
\leq 1 \}$.  $\mathbb{Z}_p$ is ring of $p$-adic integers and they
have $\nu \geq 0$ in (1). The set $\mathbb{A}_{\mathbb{Q}}$ of all
completions of $\mathbb{Q}$ in the form of the above adeles  can
be presented as
\begin{align} \label{equ.3.2}
\mathbb{A}_{\mathbb{Q}} = \bigcup_{\mathcal{P}} A (\mathcal{P})\,,
\quad A (\mathcal{P}) = \mathbb{R}\times \prod_{p\in \mathcal{P}}
\mathbb{Q}_p \times \prod_{p \not\in \mathcal{P}} \mathbb{Z}_p \,.
 \end{align} Elements of $\mathbb{A}_{\mathbb{Q}}$ satisfy componentwise addition and
multiplication and form the adele ring.

The multiplicative group of ideles $\mathbb{A}_{\mathbb{Q}}^\times$
is a subset of $\mathbb{A}_{\mathbb{Q}}$ with elements $\eta = (\eta
_\infty\,, \eta _2\,, \eta _3 \,, \cdots , \eta _p\,, \cdots)$ ,
where $\eta _\infty \in \mathbb{R}^\times = {\mathbb R} \setminus \{
0\}$ and $\eta _p \in \mathbb{Q}^\times_p = {\mathbb Q}_p \setminus
\{0 \}$ with the restriction that for all but a finite set $\mathcal
P$ one has that $\eta _p \in {\mathbb U}_p = \{ x \in \mathbb{Q}_p\,
: |x|_p = 1 \},$ i.e. ${\mathbb U}_p$ is  multiplicative group of
$p$-adic units. The entire set of adeles, related to
$\mathbb{Q}^\times = \mathbb{Q} \setminus \{0 \},$ is
\begin{align} \label{equ.3.3}
 {\mathbb A}_{\mathbb{Q}}^\times = \bigcup_{{\mathcal P}} { A}^\times ({\mathcal P}),
 \ \ \ \  A^\times ({\mathcal P}) = {\mathbb R}^{\times}\times \prod_{p\in {\mathcal P}}
 {\mathbb Q}^\times_p
 \times \prod_{p\not\in {\mathcal P}} {\mathbb U}_p \, .
\end{align}

A principal adele (idele) is a sequence $ (x, x, \cdots, x,
\cdots) \in \mathbb{A}_{\mathbb{Q}}$ , where $x \in
\mathbb{Q}\quad (x \in {\mathbb Q}^\times)$. ${\mathbb Q}$ and
${\mathbb Q}^\times$ are naturally embedded in
$\mathbb{A}_{\mathbb{Q}}$ and $\mathbb{A}_{\mathbb{Q}}^\times$ ,
respectively. This concept of  principal adeles gives way to
present rational numbers together with their nontrivial norms.
Adeles are a generalization of principal adeles
 so that it takes  into account all completions of $\mathbb{Q}$ and
 has well-defined mathematical structure.

Space of adeles (ideles) has its adelic (idelic) topology. With
respect to their topology $\mathbb{ A}_{\mathbb{Q}}$ and ${\mathbb
A}_{\mathbb{Q}}^\times$ are locally compact topological spaces.
There are adelic-valued  and complex-valued functions of adelic
arguments.  For various mathematical aspects of adeles and their
functions we refer to books \cite{gelfand,weil} and for their
applications in mathematical physics to
\cite{freund,vladimirov,dragovich3}.

Ideles and adeles are introduced in the 1930s by Claude Chevalley
and Andr\'e Weil, respectively. $p$-Adic numbers and adeles have
many applications in mathematics.  Since 1987, they have employed
in  {\it $p$-adic mathematical physics}.

Adelic connection of $p$-adic and real properties of the same
rational quantity can be well illustrated by the following two
simple examples:
\begin{align}\label{equ.3.4}
& |x|_\infty \times \prod_{p\in \mathbb{P}} |x|_p = 1\,, \,\,
\mbox{if} \,\, x \in \mathbb{Q}^\times  \\
& \chi_\infty (x) \times \prod_{p\in \mathbb{P}} \chi_p (x) = 1
\,, \,\, \mbox{if} \,\, x \in \mathbb{Q}, \label{equ.3.5}
\end{align}
where $\mathbb{P}$ is set of all primes and
\begin{align} \label{equ.3.6}
\chi_\infty (x) = \exp ( -{2 \pi i x}) , \quad \, \chi_p (x) =
\exp (2 \pi i \{x\}_p)
\end{align}
with $\{x\}_p$ as fractional part of $x$ in expansion with respect
to base $p$.

More complex connection, but also very significant, is the
Freund-Witten product formula for string amplitudes
\cite{freund3}:
\begin{align} \label{equ.3.7}
   A (a, b) =
 A_\infty (a, b) \prod_p A_p (a, b) = g_\infty^2 \,\prod_p g_p^2 = const.
\end{align}
which connects $p$-adic Veneziano amplitudes (\ref{equ.3}) with
their real analog (\ref{equ.2}). Formula (\ref{equ.3.7}) follows
as a consequence of the Euler product formula for the Riemann zeta
function applied to $p$-adic string amplitudes (\ref{equ.3}). Main
significance of (\ref{equ.3.7}) is in the fact that scattering
amplitude for real string  $A_\infty (a, b)$, which is a special
function, can be presented as product of inverse $p$-adic
amplitudes, which are elementary functions. Also, this product
formula treats $p$-adic and ordinary strings at the equal footing.
It gives rise to suppose that if there exists an ordinary scalar
string then it should exist also its $p$-adic analog. Moreover,
$p$-adic strings seem to be simpler for theoretical investigation
and useful for cosmological investigations.

\subsection{Some Adelic Cosmological Investigations}

The first consideration of $p$-adic gravity and adelic quantum
cosmology was in \cite{arefeva}. It was introduced an idea of the
fluctuating number fields at the Planck scale giving rise to
$p$-adic valued as well as  real valued gravity. Using
Hartle-Hawking approach, it was shown that the wave function for
the de Sitter minisuperspace model can be presented as an infinite
product of its $p$-adic counterparts.

Since adelic generalization of the Hartle-Hawking proposal was
serious problems in minisuperspace models with matter, further
developments of adelic quantum cosmology  were done (see
\cite{dragovich4} and references therein) using formalism of
adelic quantum mechanics \cite{dragovich5}. It was shown that
$p$-adic effects in adelic approach yield some discreteness of the
minisuperspace and cosmological constant.

Possibility that the universe is composed of real and some
$p$-adic worlds was considered in \cite{dragovich6}. In the
present paper we adopted approach that $p$-adic worlds are made of
non-tachyonic $p$-adic matter.

Let us also mention research  on $p$-adic inflation
\cite{barnaby}, and investigation of nonlocal cosmology with
tachyon condensation by rolling tachyon from a false local vacuum
to a stable one (see, e.g.,
\cite{arefeva1,arefeva2,calcagni,dragovich7} and references
therein).

\section{Concluding Remarks}
\label{sec:4}

In the present article we have introduced a non-tachyonic $p$-adic
matter which has origin in open scalar $p$-adic strings. Formally
the corresponding Lagrangian was obtained replacing $m^2$ by $-
m^2$ in Lagrangian for $p$-adic string. In space-time dimensions
$D = 2 + 4 k$ the kinetic energy term has correct sign and stable
negative local vacua. For this case there is decreasing time
dependent  field solution of the equation of motion and negative
cosmological constant.

This $p$-adic matter interacts with ordinary matter by gravity and
should play some role in the dark side of the universe. In
particular, the negative cosmological constant can change
expansion to contraction and provide bouncing in cyclic universe
evolution. These cosmological aspects are under consideration. If
 $p$-adic matter would be produced at the LHC experiment in CERN,
then its first signature should be in the form of  missing mass
(energy) in the final state, because it interacts with ordinary
matter only through gravitational interaction.

In the case of gravity with
Friedmann-Lema\^{\i}tre-Robertson-Walker (FLRW) metric the
d'Alembertian is $\Box = - \partial_t^2 - 3 H
\partial_t,$ where $H$ is the Hubble parameter $H(t) =
\frac{\dot{a}}{a}. $ Then equation of motion  contains this
operator $\Box$ and time dependent solution (\ref{equ.8n1}) for a
constant H is
\begin{align} \nonumber
 \phi_q (t) & =  q^{\frac{1}{2(q-1)}}\, \exp\Big(- \frac{3 \ln q}{2} \frac{H}{m^2} \partial_t \Big) \, \exp \Big( - \frac{q-1}{2
 \, q \ln q}\, m^2\, t^2 \Big)\\ & = q^{\frac{1}{2(q-1)}}\,  \exp \Big( - \frac{q-1}{2
 \, q \ln q}\, m^2\, \big(t - \frac{3 \ln q}{2} \frac{H}{m^2}\big)^2 \Big).  \label{equ.8n2}
\end{align}

Note that equation of motion (14) can be formally obtained from
(7) by partial replacement $p \to \frac{1}{p}$ in the following
two ways. (\textit{i}) In the LHS of (7) replace $p$ by
$\frac{1}{p}$ and $\varphi$  by $\phi$. (\textit{ii}) In the RHS
of (7) replace $p$ by $\frac{1}{p}$ and $\varphi$  by $\phi^p$.
In \cite{arefeva3}, the equation
$$
e^{-\beta \Box} \Phi(x) = \sqrt{k} \Phi^k , \quad 0 < k <1 , \,\,
(\beta > 0)
$$
was considered and it corresponds to the case (\textit{ii}) when
$k = \frac{1}{p}$.

We have also emphasized that results of measurements are rational
numbers with norm in the form of the familiar absolute value, i.e.
they are real rational numbers and not $p$-adic ones. In
Lagrangian we have made some connection between nonlocality and
$p$-adic valued world-sheet.

\section*{Acknowledgements} This investigation is supported by
Ministry of Education and Science of the Republic of Serbia, grant
No 174012. This article is based on the author talk presented at
the IX International Workshop ``Lie Theory and its Applications in
Physics'', 20--26 June 2011, Varna, Bulgaria and the author thanks
organizers for hospitality and creative scientific atmosphere. The
author also thanks I. Ya Aref'eva for useful discussions.
%\end{acknowledgement}


\begin{thebibliography}{99.}

\bibitem{kinezi} Miao Li, Xiao-Dong Li, Shuang Wang and Yi Wang, ``Dark energy'', Commun.Theor.Phys.  \textbf{56} (2011) 525-604
                  [arXiv:1103.5870v6 [astro-ph.CO]].

\bibitem{clifton} T. Clifton, P.G. Ferreira, A. Padilla, C. Skordis, ``Modified gravity and cosmology'', Phys. Rep. \textbf{513} (1) (2012) 1-189
                  [arXiv:1106.2476v2 [astro-ph.CO]].

\bibitem{freund} L. Brekke  and P.G.O. Freund, ``$p$-Adic numbers in physics'', Phys. Rep. \textbf{233} (1) (1993) 1--66.

\bibitem{vladimirov} V.S. Vladimirov, I.V. Volovich and E.I. Zelenov, \textit{$p$-Adic Analysis and Mathematical Physics}
                     (World Scientific, Singapore, 1994).

\bibitem{dragovich1} I. Dimitrijevic, B. Dragovich, J. Grujic and Z. Rakic, ``On modified gravity'', [arXiv:1202.2352v1 [hep-th]].


\bibitem{arefeva01} I.Ya. Aref'eva, ``Nonlocal string tachyon as a model for cosmological dark energy'', AIP Conf. Proc. \textbf{826} (2006)
301--311  [arXiv:astro-ph/0410443v2].

\bibitem{koshelev1} A.S. Koshelev ans S.Yu. Vernov, `` Analysis of scalar perturbations in cosmological models
                    with a non-local scalar field'',  Class. Quant. Grav. \textbf{28} (2011) 085019  [arXiv:1009.0746v2 [hep-th]].


\bibitem{volovich1} I.V. Volovich, ``$p$-Adic string'',  Class. Quant. Grav. \textbf{4} (1987)  L83--L87.

\bibitem{freund1} P.G.O. Freund and M. Olson, ``Non-archimedean strings'', Phys. Lett. B \textbf{199} (1987) 186--190.

\bibitem{gelfand} I. M. Gelf'and, M. I. Graev and I. I. Pyatetskii-Shapiro, \textit{Representation Theory and Automorphic Functions}
            (Saunders, Philadelphia, i969).

\bibitem{schikhof} W. Schikhof, \textit{Ultrametric Calculus} (Cambridge University Press, Cambridge, 1984).

\bibitem{dragovich2} B. Dragovich, A.Yu. Khrennikov, S.V. Kozyrev and I.V. Volovich, ``On $p$-adic mathematical physics'',
          $p$-Adic Numb. Ultr. Anal. Appl. \textbf{1} (1) (2009) 1--17.

\bibitem{freund2} L. Brekke, P.G.O. Freund, M. Olson and E. Witten, ``Nonarchimedean string dynamics'', Nucl. Phys. B  \textbf{302} (3)
(1988) 365--402.

\bibitem{frampton} P.H. Frampton and Y. Okada, ``Effective  scalar field theory  of $p$-adic string'',  Phys. Rev. D \textbf{37} (1988)
        3077--3084.

\bibitem{sen} D. Ghoshal and A. Sen, ``Tachyon condensation and brane descent relations in $p$-adic string theory'', Nucl. Phys.
B \textbf{584} (2000) 300-312 [arXiv:hep-th/0003278v1].

\bibitem{moeller} N. Moeller and B. Zwiebach, ``Dynamics with infinitely many time derivatives and rolling tachyons'', JHEP \textbf{0210} (2002) 034
[arXiv:hep-th/0207107v2].

\bibitem{vladimirov1}  V.S. Vladimirov, ``On some exact solutions in $p$-adic open-closed string theory'', $p$-Adic Numb. Ultr. Anal. Appl.
                 \textbf{4} (1) (2012) 57--63.

%\bibitem{yang} H. Yang, ``Stress tensors in $p$-adic string theory and truncated OSFT``, JHEP \textbf{0211} (2002) 007 [arXiv:hep-th/0209197v3].

\bibitem{weil}  A. Weil,  \textit{Adeles and Algebraic Groups} (Birkhauser, Basel, 1982).

\bibitem{dragovich3} B. Dragovich, ``Adeles in mathematical physics'',  [arXiv:0707.3876v1 [math-ph]].

\bibitem{freund3} P.G.O. Freund and E. Witten, ``Adelic string amplitudes'', Phys. Lett. \textbf{199} (1987) 191--194.

\bibitem{arefeva} A.Ya. Aref'eva, B. Dragovich, P.H. Frampton and I.V. Volovich, ``The wave function of the Universe and $p$-adic gravity'',
    Int. J. Mod. Phys. A \textbf{6} (1991) 4341--4358.

\bibitem{dragovich4} G. Djordjevic, B. Dragovich, Lj. Ne\v si\'c and I.V. Volovich, $p$-Adic and adelic minisuperspace quantum cosmology'',
Int. J. Mod. Phys. A \textbf{17} (10) (2002) 1413--1433
[arXiv:gr-qc/0105050v2].

\bibitem{dragovich5} B. Dragovich, Adelic harmonic oscillator, Int. J. Mod. Phys. A \textbf{10} (1995) 2349-2365 [arXiv:hep-th/0404160v1].

\bibitem{dragovich6} B. Dragovich, ``$p$-Adic and adelic cosmology: $p$-adic origin of dark energy and dark matter'',
\textit{$p$-Adic Mathematical Physics}, AIP Conference Proceedings
\textbf{826} (2006)  25--42 [arXiv:hep-th/0602044v1].

\bibitem{barnaby} N. Barnaby, T. Biswas and J.M. Cline, ``$p$-Adic inflation'',  JHEP \textbf{0704} (2007) 056 [arXiv:hep-th/0612230v1].

\bibitem{arefeva1}  I.Ya. Aref'eva, L.V. Joukovskaya and S.Yu. Vernov, ``Bouncing and accelerating solutions
in nonlocal stringy models'', JHEP \textbf{0707}(2007) 087
[arXiv:hep-th/0701184v4].

\bibitem{arefeva2}  I.Ya. Aref'eva and A.S. Koshelev, ``Cosmological signature of tachyon condensation'', JHEP \textbf{0809} (2008) 068
          [arXiv:0804.3570v2 [hep-th]].

\bibitem{calcagni} G. Calcagni, M. Montobbio, G. Nardelli, ``A route to nonlocal cosmology'',
Phys. Rev. D \textbf{76} (2007) 126001 [arXiv:0705.3043v3
[hep-th]].

\bibitem{dragovich7} B. Dragovich,  ``Nonlocal dynamics of $p$-adic  strings'',
Theor. Math. Phys. \textbf{164} (3)  (2010) 1151--115
[arXiv:1011.0912v1 [hep-th]].

\bibitem{arefeva3}  I.Ya. Aref'eva and L.V. Joukovskaya,  ``Time lumps in nonlocal stringy models and cosmological applications'',
JHEP \textbf{0510}(2005) 087 [arXiv:hep-th/0504200v2].

\end{thebibliography}
\end{document}